\documentclass[article,prl,balancelastpage,twocolumn,showpacs]{revtex4}
\usepackage{makeidx}
\usepackage{amssymb}
\usepackage{dcolumn}
\usepackage{graphicx}
\usepackage{acronym}

\input{tcilatex}

\begin{document}

\title{Alternative measurements of the fermion $g$-factor in the field of a
traveling circularly polarized electromagnetic wave}
\author{B. V. Gisin }
\affiliation{IPO, Ha-Tannaim St. 9, Tel-Aviv 69209, Israel. E-mail: gisin@eng.tau.ac.il}
\date{\today }

\begin{abstract}
\noindent The field of a traveling circularly polarized electromagnetic wave
and a constant magnetic field localizes fermions perpendicularly to
propagation of the wave in the cross section of the order of the wavelength.
Unusual exact solutions of the Dirac equation correspond to this
localization. Except to routine use of thin fermion beams it can be suitable
for alternative measurements of the $g$ - factor. Details and peculiarities
of the solutions in application to the measurements are considered in the
paper.
\end{abstract}

\pacs{03.65.Pm, 03.65.Ta, 31.30.jx, 06.20.Jr\hspace{20cm}}
\maketitle

\section{Introduction}

Measurements of the magnetic moment began from works of I. Rabi, who
developed the resonance radio-frequency techniques before World War II. The
technique has been significantly refined with the advent of quantum
electrodynamics \cite{quant}.

The use of a rotating coordinate system to solve magnetic resonance problems
is described in \cite{rabi}. The time-dependent eigenvalue problem may turn
into a stationary with help of such a system.

This approach makes it possible to find an exact localized solution of Dirac
equation in the field of a traveling circularly polarized electromagnetic
wave and constant magnetic field \cite{BVG}.

Temporal behavior of spin in such an electromagnetic field differs
principally from that in the standard magnetic resonance with rotating and
constant magnetic field described by the Pauli equation. This opens the way
for an alternative measurement of the fermion $g$-factor.

The present paper summarizes the peculiarities of the solutions applicable
for such measurements.\ \ \ \ \ \ \ \ \ \ \ \ \ \ \ \ \ \ \ \ \ \ \ \ \ \ \
\ \ \ \ \ \ \ \ \ \ \ \ \ \ \ \ \ \ \ \ \ \ \ \ \ \ \ \ \ \ \ \ \ \ \ \ \ \
\ \ \ \ \ \ \ \ \ \ \ \ \ \ \ \ \ \ \ \ \ 

\subsection{The Dirac equation and potential}

Consider Dirac's equation

\begin{equation}
\{-i\frac{\partial }{\partial t}-i\mathbf{\alpha }\frac{\partial }{\partial 
\mathbf{x}}-\mathbf{\alpha \mathbf{A}}+\beta \}\Psi =0  \label{Dir}
\end{equation}%
in the electromagnetic field with the potential 
\begin{eqnarray}
A_{x} &=&-\frac{1}{2}H_{z}y+\frac{1}{\Omega }H\cos (\Omega t-\Omega z),
\label{Ax} \\
A_{y} &=&\frac{1}{2}H_{z}x+\frac{1}{\Omega }H\sin (\Omega t-\Omega z).
\label{Ay}
\end{eqnarray}%
$H_{z}$ is the constant component of the magnetic field along the $z$-axis, $%
H$ is the amplitude of the circularly polarized component.

We use normalized dimensionless variables and parameters: $%
(ct,x,y,z)\rightarrow (ct,x,y,z)/\lambdabar ,$%
\begin{eqnarray}
\mathbf{A} &\mathbf{\rightarrow }&\frac{e\lambdabar }{c\hbar }\mathbf{A,}%
\text{ \ }\frac{e\lambdabar ^{2}}{c\hbar }\mathbf{H}\rightarrow \mathbf{H,}%
\text{ \ }\Omega \rightarrow \Omega \frac{\lambdabar }{c}=2\pi \frac{%
\lambdabar }{\lambda },  \label{AH} \\
E &\rightarrow &\frac{1}{mc^{2}}E,\text{ \ }p\rightarrow \frac{1}{mc}p,\text{
\ }d\rightarrow \lambdabar ^{2}d,\text{ \ }d_{2}\rightarrow \lambdabar d_{2},
\label{ndd2}
\end{eqnarray}%
$\lambdabar $ is the Compton wavelength $\lambdabar =\hbar /mc$, $\lambda $
is the wavelength of the electromagnetic wave, $c$ is the speed of light, $%
\hbar $ is the reduced Planck constant. The charge $e$, for definiteness, is
assumed to be negative $e=-|e|$. In the dimensionless units the propagation
constant equals $\Omega .$

The potential $\mathbf{A}$ describes a traveling circularly polarized
electromagnetic wave propagating along the constant magnetic field $H_{z}$.

\section{Solutions}

The Dirac's equation (\ref{Dir}) has exact solutions localized in the cross
section perpendicular to the propagation direction of the wave \cite{BVG}.
In the lab frame only non-stationary states are possible. In contrast to
that in the rotating frame stationary states exist. The solutions in the lab
frame can be presented as follows%
\begin{equation}
\Psi =\exp [-iEt+ipz-\frac{1}{2}\alpha _{1}\alpha _{2}(\Omega t-\Omega
z)+D]\psi ,  \label{psi}
\end{equation}%
where constants $E$ and $p$ are "energy" and "momentum" in the rotating
coordinate system,%
\begin{equation}
D=-\frac{d}{2}r^{2}-id_{2}\tilde{x}+d_{2}\tilde{y},  \label{D}
\end{equation}%
parameters $d,$\ $d_{2}$ are determined by substitution the wave function (%
\ref{psi}) in Eq. (\ref{Dir}) 
\begin{equation}
\text{ }d=-\frac{1}{2}H_{z},\text{\ }d_{2}=\frac{\mathcal{E}_{0}h}{2(%
\mathcal{E}-\mathcal{E}_{0})},\text{ \ }h=\frac{1}{\Omega }H,  \label{dd2}
\end{equation}%
\begin{equation}
\tilde{x}=r\cos \tilde{\varphi},\text{ \ }\tilde{y}=r\sin \tilde{\varphi},%
\text{ \ }\tilde{\varphi}=\varphi -\Omega t+\Omega z.  \label{xy}
\end{equation}

The constant spinor $\psi $ described the "ground state", spinor polynomials
in $\tilde{x}$ and $\tilde{y}$ correspond to the "excited states".

Below only the ground state is considered. In this case the wave function
has the form 
\begin{equation}
\psi =N\left( 
\begin{array}{c}
h\mathcal{E} \\ 
-(\mathcal{E}+1)(\mathcal{E}-\mathcal{E}_{0}) \\ 
h\mathcal{E} \\ 
-(\mathcal{E}-1)(\mathcal{E}-\mathcal{E}_{0})%
\end{array}%
\right) ,  \label{sol}
\end{equation}%
$N$ is a normalization constant, determined by the normalization integral $%
\int \Psi ^{\ast }\Psi ds=1,$%
\begin{equation}
\text{ }N^{2}[h^{2}\mathcal{E}^{2}+(\mathcal{E}^{2}+1)(\mathcal{E}-\mathcal{E%
}_{0})^{2}]\frac{\pi }{d}\exp (\frac{d_{2}^{2}}{d})=1,  \label{N}
\end{equation}%
$E$ obeys the characteristic equation%
\begin{equation}
\mathcal{E}(\mathcal{E}+2p-\Omega )-1-\frac{\mathcal{E}h^{2}}{\mathcal{E}-%
\mathcal{E}_{0}}=0,  \label{Eqch}
\end{equation}%
\begin{equation}
\mathcal{E}=E-cp,\text{ }\mathcal{E}_{0}=\frac{2d}{\Omega }.  \label{E0}
\end{equation}

It is noteworthy that Eq. (\ref{Eqch}) is algebraic equation of the third
order, in contrast to the even order in many other cases.

\subsection{$g$ - factor}

The parameter $\mathcal{E}_{0}$ in the non-normalized units is defined as%
\begin{equation}
\mathcal{E}_{0}=\frac{2\mu H_{z}}{\hbar \Omega },  \label{EG}
\end{equation}%
where $\mu =|e|\hbar /(2mc)$ is the Bohr magneton. The definition (\ref{EG})
by equating $2/\mathcal{E}_{0}$ to the $g$ - factor turns into the classical
condition of the magnetic resonance%
\begin{equation}
\hbar \Omega =g\mu H_{z}.  \label{cmr}
\end{equation}

This condition follows from the Pauli equation. This equation is the
nonrelativistic approximation of the Dirac equation at $g/2=1$. As it is
known from experiments $g/2$ differs from $1.$ This fact has been explicitly
explained by quantum field theory.

In the experiment $g$ - factor is calculated by means of values $H_{z}$ and $%
\Omega $ at a\ maximum amplitude of the spin flip-flop in the magnetic
resonance.

Phenomenologically this result is inserted in the Pauli equation by
multiplication of the magnetic energy $\mu \mathbf{\sigma H}$ by the
experimental value of the $g$ - factor.

Unique feature of solution (\ref{sol}) is that this solution never can be
presented as large and small two-component spinor, i. e., always corresponds
to a relativistic case. An equivalent of inserting $g$ - factor in the Pauli
equation is the equating $2/\mathcal{E}_{0}$ to the $g$ - factor in the
solution (\ref{sol}) of the Dirac equation.

\subsection{Singular solutions}

Evaluate the normalized parameter $h.$ Typically, the amplitude of the
magnetic field $H$ is much smaller than the constant magnetic field $|H_{z}|$
\begin{equation}
h=\frac{1}{\Omega }H\ll \frac{1}{\Omega }|H_{z}|\text{ }=\mathcal{E}_{0}\sim 
\frac{2}{g},  \label{h}
\end{equation}%
If $g$ $\sim 2$ then $h\ll 1.$ Therefore $h$ is very small.

Typically $\mathcal{E}$\ is expanded in power series in $h^{2}.$ However,
there exists a pair solutions for which $\mathcal{E}$\ is expanded in power
series in $h$ 
\begin{equation}
\mathcal{E=E}_{0}+h\mathcal{E}_{1}+h^{2}\mathcal{E}_{2}+\ldots ,\text{ \ }%
\mathcal{E}_{1}=\pm \frac{\mathcal{E}_{0}}{\sqrt{\mathcal{E}_{0}^{2}+1}}.
\label{Eh}
\end{equation}%
The solutions are singular solutions. Two solutions in the pair differ by
the sign of $\mathcal{E}_{k}$ with odd numbers.

The singular solutions are best suited to experiment.

For such a pair in the first approximation%
\begin{equation}
d_{2}\approx \pm \frac{\sqrt{\mathcal{E}_{0}^{2}+1}}{2}.  \label{dpm}
\end{equation}

The necessary condition for existence of the expansion (\ref{Eh}) is
following expression for the parameter $p$ (momentum in the rotating frame)
for both the states in the pair.%
\begin{equation}
p=\frac{1}{2}(\frac{1}{\mathcal{E}_{0}}-\mathcal{E}_{0})+\frac{\Omega }{2}.
\label{p}
\end{equation}%
With this momentum $p$ the energy $E$ also coincide but with accuracy $\sim
h $ 
\begin{equation}
E=\mathcal{E}_{0}+p=\frac{1}{2}(\frac{1}{\mathcal{E}_{0}}+\mathcal{E}_{0})+%
\frac{1}{2}\Omega +\ldots  \label{E}
\end{equation}%
$\Omega \ll \mathcal{E}_{0}$ (in the non-normalized units this inequality is 
$\hbar \Omega \ll mc^{2}\mathcal{E}_{0}$) and the last term in (\ref{p}) and
(\ref{E}) can be neglected.

In particular, the energy and momentum of electron in rotating frame in
non-normalized units is $E\approx mc^{2}$ and $p\approx -$ $1.165\cdot
10^{-3}mc,$ because $g/2$ differs from one by $1.165\cdot 10^{-3}$.\ 

Note that similar pairs also possible for the Pauli equation in rotating
magnetic field. Sum and difference of wave functions of the pair produce
states with the spin oscillations. Such states take part in the magnetic
resonance \cite{BVG}.

However, in the given case the states have a vanishingly small amplitude.
This amplitude is proportional the factor $\exp (-2d_{2}^{2}/d).$ In the
non-normalized units 
\begin{equation}
\frac{2d_{2}^{2}}{d}\rightarrow \frac{\mathcal{E}_{0}^{2}+1}{\mathcal{E}_{0}}%
\frac{\lambda }{2\pi \lambdabar }.  \label{ww}
\end{equation}
The typical frequency of the magnetic resonance $\sim 100GHz,$ the
wavelength corresponding to the frequency $\lambda \sim 0.3$ $cm$. The ratio 
$\lambda $ to Compton wavelength is of the order of $10^{9}.$ Therefore the
term $\exp (-2d_{2}^{2}/d)$ is extremely small and this case is not
considered in the paper.

\section{Spatial averaging}

For non-stationary solutions average values of operators should be used: $%
\overline{P}=\int \Psi ^{\ast }P\Psi dxdy$. Here and below all integrations
in respect to $x$and $y$ are from $-\infty $ to $+\infty $ and small terms
of the order of $h$ and of the order of $\hbar \Omega \ll mc^{2}$ are
neglected. Results are presented in non-normalized units.

\subsection{Localization}

The wave function (\ref{sol}) is one more example of "optimal localization"
like the ground state of the harmonic oscillator. It means the equality in
the uncertainty relation 
\begin{equation}
\overline{(\Delta x)^{2}}\overline{(\Delta p_{x})^{2}}=\frac{\hbar }{2},%
\text{ \ }\overline{(\Delta y)^{2}}\overline{(\Delta p_{y})^{2}}=\frac{\hbar 
}{2}.  \label{pn}
\end{equation}%
The parameter of the localization may be determined as follows 
\begin{equation}
2\sqrt{\int \Psi ^{\ast }(x^{2}+y^{2})\Psi dxdy}=\frac{\lambda }{\pi }\sqrt{%
1+(\frac{g}{2})^{2}}.  \label{rav}
\end{equation}%
For electron $g\sim 2$ and this parameter equals $\lambda \cdot 0.45016$

\subsection{Energy and momentum}

Neglecting terms with $h\ll 1$ and $\hbar \Omega \ll $ $mc^{2}$ obtain
energy 
\begin{equation}
\overline{E}\approx mc^{2}(\frac{g}{2}+\frac{2}{g})  \label{Eav}
\end{equation}%
and components of momentum%
\begin{eqnarray}
\overline{p_{x}} &\approx &\mp mc\frac{\sqrt{4+g^{2}}}{2g}\cos (\Omega
t-\Omega z),  \label{pa1} \\
\overline{p_{y}} &\approx &\mp mc\frac{\sqrt{4+g^{2}}}{2g}\sin (\Omega
t-\Omega z),  \label{pa2} \\
\overline{p_{z}} &\approx &mc\frac{g}{2}.  \label{pa3}
\end{eqnarray}

For electron $\overline{E}\approx 2mc^{2},$ $\overline{p_{z}}\approx mc.$

\subsection{Spin}

The average value of spin components are $s_{n}=-\frac{i}{2}\hbar \int \Psi
^{\ast }\sigma _{n}\Psi ds,$ where 
\begin{equation}
\sigma _{1}=\alpha _{2}\alpha _{3},\text{ }\sigma _{2}=\alpha _{3}\alpha
_{1},\text{ }\sigma _{3}=\alpha _{1}\alpha _{2}.  \label{sig}
\end{equation}%
In the same conditions $h\ll 1$ and $\hbar \Omega \ll $ $mc^{2}$ obtain 
\begin{eqnarray}
s_{1} &\approx &\mp \frac{\hbar g}{2\sqrt{4+g^{2}}}\cos (\Omega t-\Omega z),
\label{s1} \\
\text{\ }s_{2} &\approx &\mp \frac{\hbar g}{2\sqrt{4+g^{2}}}\sin (\Omega
t-\Omega z)  \label{s2} \\
s_{3} &\approx &0.  \label{s3}
\end{eqnarray}%
It is noteworthy that%
\begin{equation}
\frac{s_{1}}{\overline{p_{x}}}=\frac{s_{2}}{\overline{p_{y}}}=\lambdabar 
\frac{g^{2}}{(4+g^{2})}.  \label{sp}
\end{equation}%
For electron the right part of this equation $\approx \lambdabar /2.$

\subsection{Conclusion}

With help of the traveling circularly polarized electromagnetic wave and
constant magnetic field fermion can be localized in the cross section of the
order of the wavelength. Fermion states in the are described by the singular
solutions of the Dirac equation and must have a certain energy and momentum,
moreover the classical condition of the magnetic resonance should be
fulfilled.

Measurements of the $g$ - factor are possible by varying the ratio $H_{z}$
to $\Omega .$ This results to the change of the spin amplitude as well as
energy and momentum at output of the electromagnetic wave area.

Undoubtedly, accuracy of such measurements is less than that in traditional
measurements, however, such measurements are of importance because it can
give an answer on the question: is the anomalous magnetic moment a intrinsic
characteristic of fermion or it is a subject of the interpretation of
measurements?


\begin{thebibliography}{9}
\bibitem{quant} J. Schwinger, Phys. Rev. \textbf{73}, 416 (1948); \textbf{\ }%
S. S. Schweber, \emph{QED and Men Who Made It: Dyson, Feyman Schwinger, and
Tomonaga,} (Princeton University Press, Princeton, New Jersey, 1994).

\bibitem{rabi} I. I. Rabi, N. F. Ramsey and J. Schwinger, \emph{Use of
Rotating Coordinates in Magnetic Resonance Problems,} Rev. Mod. Phys. 
\textbf{26}, Iss. 2, 167\ (1954).

\bibitem{BVG} B. V. Gisin, \emph{Quantum states in rotating electromagnetic
fields,} arXiv:1011.2622v3 [math-ph], 2 Jun 2011.
\end{thebibliography}
\end{document}